\documentclass[envcountsame]{llncs}

\usepackage{t1enc} 
\usepackage{amsmath}
\usepackage{latexsym}
\usepackage{theorem}
\usepackage{amssymb}
\usepackage{url}
\usepackage{wasysym} 

\newcommand{\ifshort}[1]{\ifthenelse{\equal{\version}{short}}{#1}{}}
\newcommand{\iflong}[1]{\ifthenelse{\equal{\version}{long}}{#1}{}}

\newcommand{\comment}[1]{}
\newcommand{\bu}{$\bullet$}
\newcommand{\hs}[1][3ex]{\hspace*{#1}}

\newcommand{\vide}{\emptyset}

\newcommand{\dom}{\mr{dom}}

\newcommand{\FV}{\mr{FV}}
\newcommand{\pos}{\mr{Pos}}

\renewcommand{\a}{\rightarrow}
\newcommand{\A}{\Rightarrow}

\newcommand{\ad}{\downarrow}

\renewcommand{\to}{\mapsto}

\newcommand{\ab}{\a_\b}

\renewcommand{\ae}{\a_\eta}
\newcommand{\abe}{\a_{\b\eta}}

\newcommand{\ar}{\a_\cR}
\newcommand{\abr}{\a_{\b\cR}}

\newcommand{\al}[1][]{{}_{#1}\!\!\leftarrow}
\newcommand{\als}[1][]{{}_{#1}^*\!\!\leftarrow}

\newcommand{\I}[1]{[\![#1]\!]}

\newcommand{\et}{\wedge}

\newcommand{\st}{\star}
\newcommand{\B}{\Box} 
\renewcommand{\th}{\vdash}

\newcommand{\sle}{\subseteq}

\newcommand{\tlt}{\lhd}
\newcommand{\tgt}{\rhd}

\newcommand{\lex}{_\mr{lex}}
\newcommand{\mul}{_\mr{mul}}

\renewcommand{\o}[1]{{\overline{#1}}}
\renewcommand{\u}[1]{{\underline{#1}}}

\renewcommand{\b}{\beta}
\newcommand{\g}{\gamma}
\newcommand{\G}{\Gamma}
\renewcommand{\d}{\delta}
\newcommand{\D}{\Delta}

\newcommand{\z}{\zeta}
\renewcommand{\t}{\theta}

\newcommand{\io}{\iota}
\newcommand{\ka}{\kappa}
\newcommand{\la}{\lambda}

\renewcommand{\r}{\rho}
\newcommand{\s}{\sigma}

\newcommand{\w}{\omega}

\newcommand{\mc}{\mathcal}

\newcommand{\mr}{\mathrm}

\newcommand{\cC}{\mc{C}}

\newcommand{\cE}{\mc{E}}
\newcommand{\cF}{\mc{F}}
\newcommand{\cG}{\mc{G}}
\newcommand{\cH}{\mc{H}}

\newcommand{\cJ}{\mc{J}}

\newcommand{\cN}{\mc{N}}

\newcommand{\cR}{\mc{R}}
\newcommand{\cS}{\mc{S}}
\newcommand{\cT}{\mc{T}}

\newcommand{\cX}{\mc{X}}

\newcommand{\sR}{\ms{R}}

\newcommand{\va}{\vec{a}}
\newcommand{\vb}{\vec{b}}

\newcommand{\vl}{\vec{l}}
\newcommand{\vm}{\vec{m}}

\newcommand{\vt}{\vec{t}}
\newcommand{\vu}{\vec{u}}
\newcommand{\vv}{\vec{v}}

\newcommand{\vx}{\vec{x}}
\newcommand{\vy}{\vec{y}}

\newcommand{\vS}{\vec{S}}
\newcommand{\vT}{\vec{T}}
\newcommand{\vU}{\vec{U}}

\newenvironment{rul}%
  {$\begin{array}{rcl}}%
  {\end{array}$}
  {\begin{center}\begin{rul}}%
  {\end{rul}\end{center}}

\newenvironment{rew}[1][~~\a~~]%
  {$\begin{array}{r@{#1}l}}%
  {\end{array}$}
\newenvironment{rewc}[1][~~\a~~]%
  {\begin{center}\begin{rew}[#1]}%
  {\end{rew}\end{center}}

{\theorembodyfont{\rmfamily} 

}

\newenvironment{lstgeneric}[2]
  {\begin{list}{#1}{\topsep=.5mm\itemsep=.5mm\parsep=0mm%
    \itemindent=-3ex\labelsep=1ex\labelwidth=0ex #2}}
  {\end{list}}

\newenvironment{lst}[1]
  {\begin{lstgeneric}{#1}{\itemindent=-1ex}}
  {\end{lstgeneric}}

\newenvironment{bfenumi}[1]
  {\begin{lstgeneric}{}{\usecounter{enumi}\leftmargin=7mm%
    }}
  {\end{lstgeneric}}

\newcommand{\tf}{{\tau_f}}
\newcommand{\tg}{{\tau_g}}

\newcommand{\at}{\alpha}

\newcommand{\SN}{\cS\cN}

\newcommand{\xu}{\{x\to u\}}

\newcommand{\vxt}{\{\vx\to\vt\}}

\newcommand{\vxl}{\{\vx\to\vl\}}

\begin{document}


\title{Rewriting modulo in Deduction modulo}

\author{Fr\'ed\'eric Blanqui}

\institute{Laboratoire d'Informatique de l'\'Ecole Polytechnique\\
91128 Palaiseau Cedex, France}

\maketitle

\begin{abstract}
We study the termination of rewriting modulo a set of equations in the
Calculus of Algebraic Constructions, an extension of the Calculus of
Constructions with functions and predicates defined by higher-order
rewrite rules. In a previous work, we defined general syntactic
conditions based on the notion of computability closure for ensuring
the termination of the combination of rewriting and $\b$-reduction.

\hs[3mm]
Here, we show that this result is preserved when considering rewriting
modulo a set of equations if the equivalence classes generated by
these equations are finite, the equations are linear and satisfy
general syntactic conditions also based on the notion of computability
closure. This includes equations like associativity and commutativity
and provides an original treatment of termination modulo equations.
\end{abstract}


\section{Introduction}

The Calculus of Algebraic Constructions (CAC)
\cite{blanqui01thesis,blanqui01lics} is an extension of the Calculus
of Constructions (CC) \cite{coquand88ic} with functions and predicates
defined by (higher-order) rewrite rules. CC embodies in the same
formalism Girard's polymorphic $\la$-calculus and De Bruijn's
dependent types, which allows one to formalize propositions and proofs
of (impredicative) higher-order logic. In addition, CAC allows
functions and predicates to be defined by any set of (higher-order)
rewrite rules. And, in contrast with (first-order) Natural Deduction
Modulo \cite{dowek98trtpm}, proofs are part of the terms.

Very general conditions are studied in
\cite{blanqui01thesis,blanqui02cac} for preserving the decidability of
type-checking and the logical consistency of such a system. But these
conditions do not take into account rewriting modulo equations like
associativity and commutativity (AC), which would be very useful in
proof assistants like Coq \cite{coq} since it increases automation and
decreases the size of proofs. We already used the rewriting engine of
CiME \cite{cime}, which allows rewriting modulo AC, for a prototype
implementation of CAC, and now work on a new version of Coq including
rewriting modulo AC. In this paper, we extend the conditions given in
\cite{blanqui01thesis} to deal with rewriting modulo equations.




\section{The Calculus of Algebraic Constructions}

We assume the reader familiar with typed $\la$-calculi
\cite{barendregt92book} and rewriting \cite{dershowitz90book}. The
Calculus of Algebraic Constructions (CAC) \cite{blanqui01thesis}
simply extends CC by considering a set $\cF$ of {\em symbols} and a
set $\cR$ of {\em rewrite rules}. The terms of CAC are:

\begin{center}
$t,u\in\cT ::= s ~|~ x ~|~ f ~|~ [x:t]u ~|~ tu ~|~ (x:t)u$
\end{center}

\noindent
where $s\in\cS=\{\st,\B\}$ is a {\em sort}, $x\in\cX$ a {\em
variable}, $f\in\cF$, $[x:t]u$ an {\em abstraction}, $tu$ an {\em
application}, and $(x:t)u$ a {\em dependent product}, written $t\A u$
if $x$ does not freely occur in $u$.

The sort $\st$ denotes the universe of types and propositions, and the
sort $\B$ denotes the universe of predicate types (also called {\em
kinds}). For instance, the type $nat$ of natural numbers is of type
$\st$, $\st$ itself is of type $\B$ and $nat\A\st$, the type of
predicates over $nat$, is of type $\B$.

We use bold face letters for denoting sequences of terms. For
instance, $\vt$ is the sequence $t_1\ldots t_n$ where $n=|\vt|$ is the
length of $\vt$, and $(\vx:\vT)U$ is the term $(x_1:T_1)\ldots
(x_n:T_n)U$ (we implicitly assume that $|\vx|=|\vT|=n$).

We denote by $\FV(t)$ the set of free variables of $t$, by $\dom(\t)$
the {\em domain} of a substitution $\t$, by $\pos(t)$ the set of
Dewey's positions of $t$, by $t|_p$ the subterm of $t$ at position
$p$, and by $t[u]_p$ the replacement of $t|_p$ by $u$.

Every symbol $f$ is equipped with a sort $s_f$, an {\em arity} $\at_f$
and a type $\tf$ which may be any closed term of the form $(\vx:\vT)U$
with $|\vx|=\at_f$. The terms only built from variables and
applications of the form $f\vt$ with $|\vt|=\at_f$ are {\em
algebraic}.


A {\em typing environment} $\G$ is an ordered list of type
declarations $\vx:\vT$. If $f$ is a symbol of type $\tf=(\vx:\vT)U$,
we denote by $\G_f$ the environment $\vx:\vT$.

A rule for typing symbols is added to the typing rules of CC:

\begin{center}
(symb) \quad $\cfrac{\th\tf:s_f}{\th f:\tf}$
\end{center}

A {\em rewrite rule} is a pair $l\a r$ such that (1) $l$ is algebraic,
(2) $l$ is not a variable, and (3) $\FV(r)\sle\FV(l)$. Only $l$ has to
be algebraic: $r$ may contain applications, abstractions and
products. This is a particular case of Combinatory Reduction System
(CRS) \cite{klop93tcs} which does not need {\em higher-order
  pattern-matching}.

If $\cG\sle\cF$, $\cR_\cG$ is the set of rules whose left-hand side is
headed by a symbol in $\cG$. A symbol $f$ with $\cR_{\{f\}}=\vide$ is
{\em constant}, otherwise it is (partially) {\em defined}.

A rule is {\em left-linear} (resp. {\em right-linear}) if no variable
occurs more than once in the left-hand side (resp. right-hand side). A
rule is {\em linear} if it is both left-linear and right-linear. A
rule is {\em non-duplicating} if no variable occurs more in the
right-hand side than in the left-hand side.

A term $t$ {\em $\cR$-rewrites} to a term $t'$, written $t\ar t'$, if
there exists a position $p$ in $t$, a rule $l\a r\in\cR$ and a
substitution $\s$ such that $t|_p=l\s$ and $t'=t[r\s]_p$. A term $t$
{\em $\b$-rewrites} to a term $t'$, written $t\ab t'$, if there exists
a position $p$ in $t$ such that $t|_p=([x:U]v~u)$ and
$t'=t[v\xu]_p$. Given a relation $\a$ and a term $t$, let $\a\!\!(t)=
\{t'\in\cT~|~t\a t'\}$.

\newcommand{\br}{{\b\cR}}

Finally, in CAC, $\br$-equivalent types are identified. More
precisely, in the type conversion rule of CC, $\ad_\b$ is replaced by
$\ad_\br$:

\begin{center}
(conv) \quad $\cfrac{\G\th t:T \quad T\ad_\br T' \quad \G\th
T':s}{\G\th t:T'}$
\end{center}

\noindent
where $u\ad_\br v$ iff there exists a term $w$ such that $u\abr^* w$
and $v\abr^* w$, $\abr^*$ being the reflexive and transitive closure
of $\ab\cup\ar$. This rule means that any term $t$ of type $T$ in the
environment $\G$ is also of type $T'$ if $T$ and $T'$ have a common
reduct (and $T'$ is of type some sort $s$). For instance, if $t$ is a
proof of $P(2+2)$ then $t$ is also a proof of $P(4)$ if $\cR$ contains
the following rules:

\begin{rewc}
x+0 & x\\
x+(s~y) & s~(x+y)\\
\end{rewc}

This decreases the size of proofs and increases automation as well.

A substitution $\t$ {\em preserves typing from $\G$ to $\D$}, written
$\t:\G\leadsto\D$, if, for all $x\in\dom(\G)$, $\D\th x\t:x\G\t$,
where $x\G$ is the type associated to $x$ in $\G$. Type-preserving
substitutions enjoy the following important property: if $\G\th t:T$
and $\t:\G\leadsto\D$ then $\D\th t\t:T\t$.


For ensuring the {\em subject reduction} property (preservation of
typing under reduction), every rule $f\vl\a r$ is equipped with an
environment $\G$ and a substitution $\r$ such that,\footnote{Other
conditions are necessary that we do not detail here.} if
$f:(\vx:\vT)U$ and $\g=\vxl$ then $\G\th f\vl\r:U\g\r$ and $\G\th
r:U\g\r$. The substitution $\r$ allows to eliminate non-linearities
only due to typing and thus makes rewriting more efficient and
confluence easier to prove. For instance, the concatenation on
polymorphic lists (type $list:\st\A\st$ with constructors
$nil:(A:\st)listA$ and $cons:(A:\st)A\A listA\A listA$) of type
$(A:\st)listA\A listA\A listA$ can be defined by:

\begin{rewc}
app~A~(nil~A')~l' & l'\\
app~A~(cons~A'~x~l)~l' & cons~A~x~(app~A~x~l~l')\\
app~A~(app~A'~l~l')~l'' & app~A~l~(app~A~l'~l'')\\
\end{rewc}

\noindent
with $\G=A:\st,x:A,l:listA,l':listA$ and $\r=\{A'\to A\}$. For
instance, $app~A~(nil~A')$ is not typable in $\G$ (since
$A'\notin\dom(\G)$) but becomes typable if we apply $\r$. This does
not matter since, if an instance $app~A\s~(nil~A'\s)$ is typable then
$A\s$ is convertible to $A'\s$.




\section{Rewriting Modulo}

\newcommand{\cRs}{{\cR}}
\newcommand{\as}{\a_\cRs}
\renewcommand{\ae}{\a_\cE}
\newcommand{\aes}{\sim\as}
\newcommand{\abes}{\RHD}
\newcommand{\aseb}{\LHD}
\newcommand{\er}{{\sim\!\!\cRs}}

Now, we assume given a set $\cE$ of {\em equations} $l=r$ which will
be seen as a set of {\em symmetric} rules, that is, a set such that
$l\a r\in\cE$ iff $r\a l\in\cE$. The conditions on rules imply that,
if $l=r\in\cE$, then (1) both $l$ and $r$ are algebraic, (2) both $l$
and $r$ are headed by a function symbol, (3) $l$ and $r$ have the same
(free) variables.

Examples of equations are:

\begin{center}
\begin{tabular}{r@{ = }l@{\quad}l}
$x+y$ & $y+x$ & (commutativity of +)\\
$x+(y+z)$ & $(x+y)+z$ & (associativity of +)\\
$x\times(y+z)$ & $(x\times y)+(x\times z)$ & (distributivity of $\times$)\\
$x+0$ & $x$ & (neutrality of $0$)\\
\end{tabular}
\end{center}

\begin{center}
\begin{tabular}{r@{ = }ll}
$add~A~x~(add~A'~y~S)$ & $add~A~y~(add~A'~x~S)$\\
$union~A~S~S'$ & $union~A~S'~S$\\
$union~A~S~(union~A'~S'~S'')$ & $union~A~(union~A'~S~S')~S''$\\
\end{tabular}
\end{center}

\noindent
where $set:\st\A\st$, $empty:(A:\st)setA$, $add:(A:\st)A\A setA\A
setA$ and $union:(A:\st)setA\A setA\A setA$ formalize finite sets of
elements of type $A$. Except for distributivity which is not linear,
and the equation $x+0=x$ whose equivalence classes are infinite, all
the other equations will satisfy our strong normalization
conditions. Note however that distributivity and neutrality can always
be used as rules when oriented from left to right. Hence, the word
problem for abelian groups or abelian rings for instance can be
decided by using {\em normalized rewriting} \cite{marche94lics}.

On the other hand, the following expressions are not equations since
left and right-hand sides have distinct sets of variables:

\begin{center}
\begin{tabular}{r@{ = }l@{\quad}l}
$x\times 0$ & $0$ & ($0$ is absorbing for $\times$)\\
$x+(-x)$ & $0$ & (inverse)\\
\end{tabular}
\end{center}

Let $\sim$ be the reflexive and transitive closure of $\a_\cE$ ($\sim$
is an equivalence relation since $\cE$ is symmetric). We are now
interested in the termination of $\abes=\ab\cup\aes$ (instead of
$\ab\cup\ar$ before). In the following, we may denote $\ae$ by $\cE$,
$\as$ by $\cRs$ and $\ab$ by $\b$.

In order to preserve all the basic properties of the calculus, we do
not change the shape of the relation used in the type conversion rule
(conv): two types $T$ and $T'$ are convertible if $T\ad T'$ with
$\a=\ab\cup\ar\cup\ae$. But this raises the question of how to check
this condition, knowing that $\a$ may be not terminating. We study
this problem in Section \ref{sec-confl}.




\section{Conditions of strong normalization}
\label{sec-cond}


In the strong normalization conditions, we distinguish between {\em
first-order} symbols (set $\cF_1$) and {\em higher-order} symbols (set
$\cF_\w$). To precisely define what is a first-order symbol, we need a
little definition before. We say that a constant predicate symbol is
{\em primitive} if it is not polymorphic and if its constructors have
no functional arguments. This includes in particular any first-order
data type (natural numbers, lists of natural numbers, etc.). Now, a
symbol $f$ is {\em first-order} if it is a predicate symbol of {\em
maximal arity},\footnote{A predicate symbol $f$ of type $(\vx:\vT)U$
is of {\em maximal arity} if $U=\st$, that is, if the elements of type
$f\vt$ are not functions.} or if it is a function symbol whose output
type is a primitive predicate symbol. Any other symbol is {\em
higher-order}. Let $\cR_\io= \cR_{\cF_\io}$ and $\cE_\io=
\cE_{\cF_\io}$ for $\io\in\{1,\w\}$.

Since the pioneer works on the combination of $\la$-calculus and
first-order rewriting \cite{breazu89icalp,okada89issac}, it is well
known that the addition at the object level of a strongly normalizing
first-order rewrite system preserves strong normalization. This comes
from the fact that first-order rewriting cannot create
$\b$-redexes. On the other hand, higher-order rewriting can create
$\b$-redexes. This is why we have other conditions on higher-order
symbols than merely strong normalization. Furthermore, in order for
the two systems to be combined without losing strong normalization
\cite{toyama87ipl}, we also require first-order rules to be
non-duplicating \cite{rusinowitch87ipl}. Note however that a
first-order symbol can always be considered as higher-order (but the
strong normalization conditions on higher-order symbols may not be
powerful enough for proving the termination of its defining rules).


\newcommand{\thc}{\th_\mr{\!\!c}}

The strong normalization conditions on higher-order rewrite rules are
based on the notion of {\em computability closure}
\cite{blanqui02tcs}. We are going to use this notion for the equations
too.

Typed $\la$-calculi are generally proved strongly normalizing by using
Tait and Girard's technique of {\em computability
predicates/reducibility candidates} \cite{girard88book}. Indeed, a
direct proof of strong normalization by induction on the structure of
terms does not work. The idea of Tait, later extended by Girard to the
polymorphic $\la$-calculus, is to strengthen the induction hypothesis
as follows. To every type $T$, one associates a set $\I{T}\sle\SN$
(set of strongly normalizing terms), and proves that every term of
type $T$ is {\em computable}, that is, belongs to $\I{T}$.

Now, if we extend such a calculus with rewriting, for preserving
strong normalization, a rewrite rule has to preserve
computability. The {\em computability closure} of a term $t$ is a set
of terms that are computable whenever $t$ itself is computable. So, if
the right-hand side $r$ of a rule $f\vl\a r$ belongs to the
computability closure of $\vl$, a condition called the {\em General
Schema}, then $r$ is computable whenever the terms in $\vl$ are
computable.

Formally, the computability closure for a rule $(f\vl\a r,\G,\r)$ with
$\tf=(\vx:\vT)U$ and $\g=\vxl$ is the set of terms $t$ such that the
judgment $\thc t:U\g\r$ can be deduced from the rules of Figure
\ref{fig-cc}, where the variables of $\dom(\G)$ are considered as
symbols ($\tau_x=x\G$), $>_\cF$ is a well-founded quasi-ordering
(precedence) on symbols, with $x<_\cF f$ for all $x\in\dom(\G)$, $>_f$
is the multiset or lexicographic extension\footnote{Or a simple
combination thereof, depending on the {\em status} of $f$.} of the
subterm ordering\footnote{We use a more powerful ordering for dealing
with recursive definitions on types whose constructors have functional
arguments.} $\tgt$, and $T\ad_f T'$ iff $T$ and $T'$ have a common
reduct by $\a_f=\ab\cup\a_{\cR_f^<}$ where $\cR_f^<=\{g\vu\a
v\in\cR~|~ g<_\cF f\}$.

In addition, every variable $x\in\dom(\G)$ is required to be {\em
accessible} in some $l_i$, that is, $x\s$ is computable whenever
$l_i\s$ is computable. The arguments of a constructor-headed term are
always accessible. For a function-headed term $f\vt$ with
$f:(\vx:\vT)C\vv$ and $C$ constant, only the $t_i$'s such that $C$
occurs positively in $T_i$ are accessible ($X$ occurs positively in
$Y\A X$ and negatively in $X\A Y$).

The relation $\thc$ is similar to the typing relation $\th$ of CAC
except that symbol applications are restricted to symbols smaller than
$f$, or to arguments smaller than $\vl$ in the case of an application
of a symbol equivalent to $f$. So, verifying that a rule satisfies the
General Schema amounts to check whether $r$ has type $U\g\r$ with the
previous restrictions on symbol applications. It therefore has the
same complexity.

\begin{figure}[ht]
\caption{Computability closure for $(f\vl\a r,\G,\r)$\label{fig-cc}}
\centering
\begin{tabular}{ccc}
(ax) & $\cfrac{}{\thc \st:\B}$\\\\

(symb$^<$) & $\cfrac{\thc \tg:s_g}{\thc g:\tg}$ & $(g<_\cF f)$\\\\

(symb$^=$) & $\cfrac{
\thc\tg:s_g\quad \d:\G_g\leadsto_c\D}
{\D\thc g\vy\d:V\d}$ &
$\begin{array}{c}
(\tg=(\vy:\vU)V,\\
g =_\cF f \mbox{ and } \vy\d<_f\vl)\\
\end{array}$\\\\

(var) & $\cfrac{\D\thc T:s}{\D,x:T\thc x:T}$
& $(x\notin\dom(\D))$\\\\

(weak) & $\cfrac{\D\thc T:s\quad \D\thc u:U}{\D,x:T\thc u:U}$
& $(x\notin\dom(\D))$\\\\

(abs) & $\cfrac{\D,x:U\thc v:V\quad \D\thc (x:U)V:s}
{\D\thc [x:U]v:(x:U)V}$\\\\

(app) & $\cfrac{\D\thc t:(x:U)V\quad \D\thc u:U}{\D\thc tu:V\xu}$\\\\

(prod) & $\cfrac{\D,x:U\thc V:s}{\D\thc (x:U)V:s}$\\\\

(conv) & $\cfrac{\D\thc t:T\quad \D\thc T:s\quad \D\thc T':s}{\D\thc t:T'}$
& $(T\ad_f T')$\\
\end{tabular}
\end{figure}

Now, how the computability closure can help us in dealing with
rewriting modulo equations? When one tries to prove that every term is
computable, in the case of a term $f\vt$, it is sufficient to prove
that every reduct of $f\vt$ is computable. In the case of a
head-reduct $f\vl\s\a r\s$, this follows from the fact that $r$
belongs to the computability closure of $\vl$ since, by induction
hypothesis, the terms in $\vl\s$ are computable.

Now, with rewriting modulo, a $\cRs$-step can be preceded by
$\cE$-steps: $f\vt\ae^* g\vu\as t'$. To apply the previous method with
$g\vu$, we must prove that the terms in $\vu$ are computable. This can
be achieved by assuming that the equations also satisfy the General
Schema in the following sense: an equation $(f\vl\a g\vm,\G,\r)$ with
$\tg=(\vx:\vT)U$ and $\g=\{\vx\to\vm\}$ satisfies the General Schema
if, for all $i$, $\thc m_i:T_i\g\r$, that is, the terms in $\vm$
belong to the computability closure of $\vl$. By symmetry, the terms
in $\vl$ belong to the computability closure of $\vm$.


One can easily check that this condition is satisfied by commutativity
(whatever the type of $+$ is) and associativity (if both $y$ and $z$
are accessible in $y+z$):

\begin{rewc}[~=~]
x+y & y+x\\
x+(y+z) & (x+y)+z\\
\end{rewc}

For commutativity, this is immediate and does not depend on the type
of $+$: both $y$ and $x$ belong to the computability closure of $x$
and $y$.

For associativity, we must prove that both $x+y$ and $z$ belong to the
computability closure $\cC\cC$ of $x$ and $y+z$. If we assume that
both $y$ and $z$ are accessible in $y+z$ (which is the case for
instance if $+:nat\A nat\A nat$), then $z$ belongs to $\cC\cC$ and, by
using a multiset status for comparing the arguments of $+$, $x+y$
belongs to $\cC\cC$ too since $\{x,y\}\tlt\mul\{x,y+z\}$.


We now give all the strong normalization conditions.

\newcommand{\aesun}{\sim_1\a_{\cR_1}}

\begin{theorem}[Strong normalization of $\b\cup\er$]
Let $\sim_1$ be the reflexive \hfill and transitive closure of
$\cE_1$. The relation $\abes=\ab\cup\aes$ is strongly normalizing if
the following conditions adapted from \cite{blanqui01thesis} are
satisfied:

\begin{lst}{\bu}
\item $\a=\ab\cup\as\cup\ae$ is confluent,\footnote{If there are
type-level rewrite rules.}

\item the rules of $\cR_1$ are non-duplicating,\footnote{If there are
higher-order rules.} $\cR_1\cap\cF_\w= \cE_1\cap\cF_\w=
\vide$\footnote{First-order rules/equations only contain first-order
symbols.} and $\aesun$ is strongly normalizing on first-order
algebraic terms,

\item the rules of $\cR_\w$ satisfy the General Schema and are {\em
safe},\footnote{No pattern-matching on predicates.}

\item rules on predicate symbols have no critical pair, satisfy the
General Schema\footnote{There are other possibilities. See
\cite{blanqui01thesis} for more details.} and are {\em
small},\footnote{A rule $f\vl\a r$ is {\em small} if every predicate
variable in $r$ is equal to one of the $l_i$'s.}
\end{lst}

\noindent
and if the following new conditions are satisfied too:

\begin{lst}{\bu}
\item there is no equation on predicate symbols,
\item $\cE$ is linear,
\item the equivalence classes modulo $\sim$ are finite,
\item every rule $(f\vl\a g\vm,\G,\r)\in\cE$ satisfies the General
Schema in the following sense: if $\tg= (\vx:\vT)U$ and
$\g=\{\vx\to\vm\}$ then, for all $i$, $\thc m_i:T_i\g\r$.
\end{lst}
\end{theorem}

Not allowing equations on predicate symbols is an important
limitation. However, one cannot have equations on connectors if one
wants to preserve the Curry-Howard isomorphism. For instance, with
commutativity on $\et$, one looses subject reduction. Take
$\et:\st\A\st\A\st$, $pair:(A:\st)(B:\st)A\A B\A A\et B$ and
$\pi_1:(A:\st)(B:\st)A\et B\A A$ defined by
$\pi_1~A~B~(pair~A'~B'~a~b)\a a$. Then, $\pi_1~B~A~(pair~A~B~a~b)$ is
of type $B$ but $a$ is not.




\section{Strong normalization proof}
\label{sec-sn-proof}

The strong normalization proof follows the one given in
\cite{blanqui03short} very closely.\footnote{The proof given in
\cite{blanqui03short} is an important simplification of the one given
in \cite{blanqui01thesis}.} We only give the definitions and lemmas
that must be modified. As previously explained, the strong
normalization is obtained by defining an interpretation $\I{T}\sle\SN$
for every type $T$, and by proving that every term of type $T$ belongs
to $\I{T}$.

More precisely, for every type $T$, we define the set $\cR_T$ of the
possible interpretations, or {\em candidates}, for the terms of type
$T$. $\cR_{(x:U)V}$ is the set of functions $R$ from $\cT\times\cR_U$
to $\cR_V$ that are stable by reduction: if $u\a u'$ then
$R(u,S)=R(u',S)$. A term $t$ is {\em neutral} if it is distinct from
an abstraction or a constructor. $\cR_\st$ is the set of sets
$R\sle\cT$ such that:

\begin{bfenumi}{R}
\item Strong normalization: $R\sle\SN$.
\item Stability by reduction: if $t\in R$ then $\a\!\!(t)\sle R$.
\item Neutral terms: if $t$ is neutral and $\abes(t)\sle R$ then
$t\in R$.
\end{bfenumi}

Candidates form a complete lattice. A {\em candidate assignment} $\xi$
is a function which associates a candidate to every variable. Given an
interpretation $I$ for predicate symbols, a candidate assignment $\xi$
and a substitution $\t$, the {\em interpretation} of a type $T$,
written $\I{T}^I_{\xi,\t}$, is defined in \cite{blanqui02cac}. The
elements of $\I{T}^I_{\xi,\t}$ are said {\em computable}. A pair
$(\xi,\t)$ is {\em $\G$-valid}, written $\xi,\t\models\G$, if, for all
$x\in\dom(\G)$, $x\xi\in\cR_{x\G}$ and $x\t\in \I{x\G}^I_{\xi,\t}$.

Then, strong normalization is obtained by defining an interpretation
$I_f\in\cR_\tf$ for every predicate symbol $f$, and by proving that
every symbol $f$ is computable, that is, $f\in\I{\tf}$. If
$\tf=(\vx:\vT)U$, it amounts to check that, for all $\G_f$-valid pair
$(\xi,\t)$, $f\vx\t\in \I{U}_{\xi,\t}$. For the interpretation, we
keep the one for constant predicate symbols given in
\cite{blanqui03short} but slightly modify the interpretation of
defined predicate symbols for taking into account the new reduction
relation.


Although we do not change the interpretation of constant predicate
symbols, we must check that the interpretation of {\em primitive}
predicate symbols is still $\SN$ (hence that, for primitive predicate
symbols, computability is equivalent to strong normalization), since
this property is used for proving that a terminating and
non-duplicating (if there are higher-order rewrite rules) first-order
rewrite system preserves strong normalization. The verification of the
former property is easy. We now prove the latter.


\newcommand{\etgt}{\sim\!\!\tgt}

\begin{lemma} {\bf\cite{jouannaud86siam}}\,
\label{lem-etgt}
If the $\sim$-classes are finite then $\etgt$ is strongly normalizing.
\end{lemma}

\begin{proof}
We prove that $(\etgt)^n\sle\,\etgt^n$ by induction on $n$. For $n=0$,
this is immediate. For $n+1$, $(\etgt)^{n+1}\sle\,
\etgt\!\!\etgt^n\sle\, \sim\etgt\tgt^n\sle\, \etgt^{n+1}$.\qed
\end{proof}


\newcommand{\tgte}{\tgt\!\!\sim}
\renewcommand{\abe}{\ab\sim}
\newcommand{\cRsun}{{\cR_1}}
\newcommand{\ase}{\as\sim}
\newcommand{\asun}{\a_\cRsun}
\newcommand{\aeun}{\a_{\cE_1}}
\newcommand{\aseun}{\asun\sim_1}
\newcommand{\absun}{\a_{\b\cRsun}}
\newcommand{\arp}{\twoheadrightarrow}

\begin{lemma} {\bf\cite{dougherty92ic}}\,
\label{lem-beta-run}
If $t\in\SN(\b)$ and $t\asun u$ then $\b(t)\asun^* \b(u)$.
\end{lemma}

\begin{proof}
Dougherty proves this result in \cite{dougherty92ic} (Proposition 4.6
and Theorem 4.7) for the untyped $\la$-calculus. The proof can clearly
be extended to the Calculus of Algebraic Constructions. We inductively
define $\arp$ as follows:

\begin{lst}{\bu}
\item $a\arp a$;
\item if $l\a r\in\cRsun$ and $\s\arp \t$ then $l\s\arp r\t$;
\item if $a\arp b$ and $c\arp d$ then $ac\arp bd$, $[x:a]c\arp
[x:b]d$ and $(x:a)c\arp (x:b)d$;
\item if $\va\arp \vb$ then $f\va\arp f\vb$.
\end{lst}

We now prove that, if $t\ab t'$ and $t\arp u$ then there exist $t''$
and $u'$ such that $t'\ab^* t''\arp u'$ and $u\ab^* u'$ by induction
on $t\arp u$.

\begin{lst}{\bu}
\item $u=t$. Immediate.

\item $t=l\s$, $u=r\t$ and $\s\arp \t$. Since left-hand sides of rules are
algebraic, the $\b$-reduction must take place in an occurrence of a
variable $x\in\FV(l)$. Let $v'$ be the $\b$-reduct of $x\s$. By
induction hypothesis, there exists $v''$ and $w$ such that $v'\ab^*
v''\arp w$ and $x\t\ab^* w$. Let $\s''$ such that $x\s''=v''$ and
$y\s''=y\s$ if $y\neq x$, and $\t'$ such that $x\t'=w$ and $y\t'=y\t$
if $y\neq x$. We have $\s''\arp \t'$. By $\b$-reducing all the
instances of the occurrences of $x$ in $l$ to $v''$, we get $t'\ab^*
l\s''\arp r\t'$ and, by reducing all the instances of the occurrences
of $x$ in $r$ to $w$, we get $u=r\t\ab^* r\t'$.

\item Assume that $t=[x:a]c~k$, $u=v~l$, $[x:a]c\arp v$, $k\arp l$ and
$t'=c\{x\to k\}$. Then, $v=[x:b]d$ with $a\arp b$ and $c\arp
d$. Therefore, $c\{x\to k\}\arp d\{x\to l\}$ and $u\ab d\{x\to l\}$.

Assume now that $t=ac$, $u=bd$, $a\arp b$, $c\arp d$ and $a\ab
a'$. The other cases are similar. By induction hypothesis, there exist
$a''$ and $b'$ such that $a'\ab^* a''\arp b'$ and $b\ab^*
b'$. Therefore, $a'c\ab^* a''c\arp b'd$ and $bd\ab^* b'd$.

\item $t=f\va$, $u=f\vb$ and $\va\arp \vb$. Then, there is $i$ such
that $t'=f\va'$, $a_i\ab a_i'$ and $a_j=a_j'$ if $j\neq i$. By
induction hypothesis, there exists $a_i''$ and $b_i'$ such that
$a_i'\ab^* a_i''\arp b_i'$ and $b_i\ab^* b_i'$. Let $a_j''=a_j$ and
$b_j'=b_j$ if $j\neq i$. Then, $\va''\arp \vb'$, $t'=f\va'\ab^*
f\va''\arp f\vb'$ and $u=f\vb\ab^* f\vb'$.
\end{lst}

Now, since $t$ is $\b$-strongly normalizable, we can prove the lemma
by induction on $\ab$. If $t$ is in $\b$-normal form then $u$ also is
in $\b$-normal form since $\cRsun$-reductions preserve $\b$-normal
forms. Hence, $\b(t)=t\arp u=\b(u)$. Now, if $t\ab t'$ then there
exist $t''$ and $u'$ such that $t'\ab^* t''\arp u'$ and $u\ab^*
u'$. By induction hypothesis, $\b(t'')\arp \b(u')$. Therefore,
$\b(t)\arp \b(u)$.\qed
\end{proof}


\begin{definition}[Cap and aliens]
  Let $\z$ be an injection from the classes of terms modulo $\ad^*$ to
  $\cX$. The {\em cap} of a term $t$ is the biggest first-order
  algebraic term $cap(t)= t[x_1]_{p_1} \ldots [x_n]_{p_n}$ such that
  $x_i= \z(t|_{p_i})$. The $t|_{p_i}$'s are called the {\em aliens} of
  $t$. We denote by $\b(t)$ the $\b$-normal form of $t$, by $cap\b(t)$
  the cap of $\b(t)$, by $Cap(t)$ (resp. $Cap\b(t)$) the
  $\sim_1$-equivalence class of $cap(t)$ (resp. $cap\b(t)$), by
  $aliens(t)$ the multiset of the aliens of $t$, and by $Aliens(t)$
  the multiset union of the (finite) $\sim$-equivalence classes of the
  aliens of $t$.
\end{definition}


\begin{theorem}[Computability of first-order symbols]
  If $f\in\cF_1$ and $\vt\in\SN$ then $f\vt\in\SN$.
\end{theorem}

\begin{proof}
  We prove that every $\abes$-reduct $t'$ of $t=f\vt$ is strongly
  normalizable. In the following, $(>_a,>_b)\lex$ denotes the
  lexicographic ordering built with $>_a$ and $>_b$, and $>\mul$
  denotes the multiset extension of $>$.
  
  \u{\bf Case $\cR_\w\neq\vide$}. By induction on $(Aliens(t),Cap(t))$
  with $((\abe\cup\ase\cup\,\tgte)\mul$, $(\aseun)\mul)\lex$ as
  well-founded ordering. It is easy to see that the aliens are
  strongly normalizable for $\abe$, $\ase$ and $\tgte$ since they are
  so for $\sim\ab$ (Lemma \ref{lem-beta-com}), $\aes$ and $\etgt$
  (Lemma \ref{lem-etgt}) respectively.

  If $t\ab t'$ then the reduction takes place in an alien $v$. Let
  $v'$ be its $\b$-reduct. If $v'$ is not headed by a symbol of
  $\cF_1$ then $Aliens(t) ~(\ab\sim)\mul~ Aliens(u)$. Otherwise, its
  cap increases the cap of $t'$ but, since the aliens of $t'$ are then
  strict subterms of $v'$, we have $Aliens(t)
  ~(\ab\sim\cup\,\tgte)\mul~ Aliens(u)$.

  Assume now that $t\ae^* u\as t'$. We first look at what happens when
  $t\ae u$. There are two cases:
\begin{lst}{\bu}
\item If the reduction takes place in the cap then this is a
  $\cE_1$-reduction. Since both the left-hand side and the right-hand
  side of a first-order rule are first-order algebraic terms, we have
  $cap(t)\aeun cap(u)$ and, since the rules of $\cE$ are linear, we
  have $aliens(t)=aliens(u)$.
\item If the reduction takes place in an alien then $cap(t)=cap(u)$
  and $aliens(t)$ $~(\a_\cE)\mul~$ $aliens(u)$.
\end{lst}

  So, in both cases, $Cap(t)= Cap(u)$ and $Aliens(t)=
  Aliens(u)$. Therefore, by induction on the number of $\cE$-steps, if
  $t\ae^* u$ then $Cap(t)= Cap(u)$ and $Aliens(t)= Aliens(u)$. We now
  look at the $\cRs$-reduction. There are two cases:
\begin{lst}{\bu}
\item If the reduction takes place in the cap then it is a
  $\cRsun$-reduction. Since both the left-hand side and the right-hand
  side of a first-order rule are first-order algebraic terms, we have
  $cap(u)\asun cap(t')$ and, since the rules of $\cRsun$ are
  non-duplicating, we have $aliens(u)\sle aliens(t')$. If
  $aliens(u)\subsetneq aliens(t')$ then $Aliens(u)\subsetneq
  Aliens(t')$. Otherwise, $Cap(u) ~(\aseun)\mul~ Cap(t')$.
\item If the reduction takes place in an alien then, as in the case of
  a $\b$-reduction, we have $Aliens(t)$ $~(\ase\cup\,\tgte)\mul~$
  $Aliens(u)$.
\end{lst}
  
  \u{\bf Case $\cR_\w=\vide$}. Since the $t_i$'s are strongly
  normalizable and no $\b$-reduction can take place at the top of $t$,
  $t$ has a $\b$-normal form. We prove that every $\abes$-reduct $t'$
  of $t$ is strongly normalizable, by induction on
  $(Cap\b(t),Aliens(t))$ with $((\aseun)\mul$,
  $(\abe\cup\ase\cup\,\tgte)\mul)\lex$ as well-founded ordering.

  If $t\ab t'$ then $cap\b(t)=cap\b(t')$ and, as seen in the previous
  case, $Aliens(t)$ $~(\ab\sim\cup\,\tgte)~$ $Aliens(u)$.

  Otherwise, $t\ae^* u\asun t'$. As seen in the previous case,
  $cap(t)\aeun^* cap(u)$ and $Aliens(t)=Aliens(u)$. Since $\b$ and
  $\cE$ commute and $\cE$ preserves $\b$-normal forms, we have
  $cap\b(t)\aeun^* cap\b(u)$ and thus $Cap\b(t)=Cap\b(u)$. We now look
  at the $\cRsun$-reduction. There are two cases:
\begin{lst}{\bu}
\item The reduction takes place in the cap. Since both the
  left-hand side and the right-hand side of a first-order rule are
  first-order algebraic terms, we have $cap(u)\asun cap(t')$ and,
  since $\b$-reductions cannot reduce the cap, we have $cap\b(u)\asun
  cap\b(t')$ and thus $Cap\b(t)$ $~(\aseun)\mul~$ $Cap\b(t')$.
\item If the reduction takes place in an alien then $Aliens(t)$
  $~(\ase)\mul~$ $Aliens(u)$ and, after Lemma \ref{lem-beta-run},
  $\b(u)\asun^* \b(t')$. Therefore, $cap\b(u)\asun^* cap\b(t')$ and
  $Cap\b(u) ~(\ase)\mul~ Cap\b(t')$.\qed
\end{lst}
\end{proof}


We now come to the interpretation of defined predicate symbols. Let
$f$ be a defined predicate of type $(\vx:\vT)U$. We define
$I_f(\vt,\vS)$ by induction on $\vt,\vS$ as follows. If there exists a
rule $(f\vl\a r,\G,\r)$ and a substitution $\s$ such that
$\vt~\abes^*\sim \vl\s$ and $\vl\s$ is in $\abes$-normal form, then
$I_f(\vt,\vS)=\I{r}^I_{\xi,\s}$ with $\s=\vxt$ and $x\xi=S_{\ka_x}$
where $\ka_x$ is given by smallness. Otherwise, we take the greatest
element of $\cR_U$.

We must make sure that the definition does not depend on the choice of
the rule. Assume that there is another rule $(f\vl'\a r',\G',\r')$ and
a substitution $\s'$ such that $\vt~\abes^*\sim \vl'\s'$ in normal
form. By confluence and Lemma \ref{lem-equiv}, we have $\vl\s\sim
\vl'\s'$. Since $\a$ is confluent and rules on predicate symbols have
no critical pair, there exists $\s''$ such that $\s\ae^*\s''$,
$\s'\ae^*\s''$ and $\vl\s''=\vl'\s''$. Therefore, for the same reason,
we must have $\vl=\vl'$ and $r=r'$.

Finally, we check that the interpretation is stable by reduction: if
$t\a t'$ then, since $\a$ is confluent, $t$ has a $\abes$-normal form
iff $t'$ has a $\abes$-normal form too.

We now prove the computability of higher-order symbols.

\begin{theorem}[Computability of higher-order symbols]
If $f\in\cF_\w$, $\tf=(\vx:\vT)U$ and $\xi,\t\models\G_f$ then
$f\vx\t\in \I{U}_{\xi,\t}$.
\end{theorem}

\begin{proof}
The proof follows the one given in \cite{blanqui03short} except that
$\a$ is replaced by $\abes$. We examine the different $\abes$-reducts
of $f\vx\t$. If this is a $\b$-reduction, it must take place in one
$x_i\t$ and we can conclude by induction hypothesis. Otherwise, we
have $f\vx\t\ae^* g\vu\as t'$. Since the equations satisfy the General
Schema, the $u_i$'s are computable. Now, if the $\cRs$-reduction takes
place in one $u_i$, we can conclude by induction
hypothesis. Otherwise, this is a head-$\cRs$-reduction and we can
conclude by correctness of the computability closure.\qed
\end{proof}




\section{Confluence}
\label{sec-confl}

We now study the confluence of $\a$ and the decidability of $\ad^*$.
Let $R$ be a relation. $\o{R}$, $R^+$, $R^*$ respectively denote the
inverse, the transitive closure, and the reflexive and transitive
closure of $R$. Composition is denoted by juxtaposition.

\begin{lst}{--}
\item $R$ is {\em confluent} if $\o{R}^*R^*\sle R^*\o{R}^*$.
\item $R$ is {\em confluent modulo $\sim$} or {\em
$\sim$-confluent}\footnote{The definitions of confluence modulo and
local confluence modulo are those of \cite{jouannaud86siam}. They
differ from Huet's definition \cite{huet80acm}. Huet's confluence
modulo corresponds to our confluence modulo on equivalence classes,
but Huet's local confluence modulo does not correspond to our local
confluence modulo on equivalence classes.} if $\o{R}^*R^*\sle R^*\sim
\o{R}^*$.
\item $R$ is {\em $\sim$-confluent on $\sim$-classes} if $\o{R}^*\sim
R^*\sle R^*\sim \o{R}^*$.
\item $R$ is {\em locally confluent} if $\o{R}R\sle R^*\o{R}^*$.
\item $R$ is {\em locally $\sim$-confluent} if $\o{R}R\sle R^*\sim
\o{R}^*$.
\item $R$ is {\em locally $\sim$-confluent on $\sim$-classes} if
$\o{R}\sim R\sle R^*\sim \o{R}^*$.
\item $R$ is {\em locally $\sim$-coherent} if $\cE R\sle
R^*\sim\o{R}^*$.
\item $R$ and $S$ {\em commute} if $\o{R}S\sle S\o{R}$.
\item $R$ {\em $\sim$-commutes on $\sim$-classes} if $\o{R}\sim R\sle
R\sim\o{R}$.
\end{lst}


\begin{lemma}
\label{lem-beta-com}
If $\cE$ is linear then $\sim$ commutes with $\b$ and $\abes$.
\end{lemma}

\begin{proof}
Assume that $t\a_{\b,p} u$ ($\b$-reduction at position $p$) and
$t\a_{\cE,q} v$ ($\cE$-reduction at position $q$). There are several
cases depending on the relative positions of the different reductions.

\begin{lst}{\bu}
\item $p$ and $q$ have no common prefix. Then the reductions clearly
commute and $\cE\b\sle\b\cE$ in this case (remember that $\o\cE=\cE$).

\item $p=q$: not possible since left-hand sides of rules are algebraic
and distinct from a variable.

\item $p<q$: $t|_p= [x:A]b~a$ and $u=t[b\t]_p$ with $\t=\{x\to a\}$.
\begin{lst}{--}
\item Reduction in $A$: $v= t[[x:A']b~a]_p$ with $A\ae A'$. Then,
$v\ab u$ and $\cE\b\sle\b$.
\item Reduction in $b$: $v= t[[x:A]b'~a]_p$ with $b\ae b'$. Then,
$v\ab t[b'\t]_p~\al[\cE] u$ and $\cE\b\sle\b\cE$.
\item Reduction in $a$: $v= t[[x:A]b~a']_p$ with $a\ae a'$. Let
$\t'=\{x\to a'\}$. Then, $v\ab t[b\t']_p~\als[\cE] u$ and
$\cE\b\sle\b\cE^*$.
\end{lst}

\item $p>q$: $t=t[l\s]_q$ and $v=t[r\s]_q$. Since left-hand sides of
rules are algebraic, there is one occurrence of a variable
$x\in\FV(l)$ such that $x\s\ab w$. Let $\s'$ be the substitution such
that $x\s'=w$ and $y\s'=y\s$ if $y\neq x$. Let $a$ (resp. $b$) be the
number of occurrences of $x$ in $l$ (resp. $r$). Then, $u\ab^{a-1}
t[l\s']_q\ae t[r\s']_q~_\b^b\!\!\leftarrow v$. Since $\cE$ is linear,
we have $a=b=1$ and thus $\cE\b\sle\b\cE$.
\end{lst}

In conclusion, in every case, we have $\cE\b\sle\b\cE^*$. By induction
on the number of $\cE$-steps, we get $\cE^*\b\sle\b\cE^*$, that is,
$\sim\b\sle\b\sim$. Therefore, $\sim\abes\sle\abes\sim$ since
$\abes=\b\,\cup\sim\!\cRs$, $\sim\b\sle \b\sim\,\sle \abes\sim$ and
$\sim\sim\!\cRs\sle \abes\sim$.\qed
\end{proof}


\begin{corollary}
If $\cE$ is linear and $t\in\SN(\b)$ then $t\in\SN(\sim\!\!\b)$.
\end{corollary}

\begin{proof}
Assume that $t\in\SN(\b)$. We prove that $(\sim\!\!\b)^n \sle
\b^n\!\!\sim$ by induction on $n$. For $n=0$, this is immediate. For
$n+1$, $(\sim\b)^{n+1}= (\sim\b)^n\sim\b\sle \b^n\sim\sim\b\sle
\b^{n+1}\sim$. Therefore, $t\in\SN(\sim\b)$.\qed
\end{proof}


\newcommand{\staseb}{{}^*\aseb}

\begin{lemma}
\label{lem-ad}
If $\cE$ is linear then $\a^*\sle\,\abes^*\sim$ and
$\ad\,=\,\abes^*\sim\staseb$.
\end{lemma}

\begin{proof}
$\a^*\,\sle (\b\cup\cE\cup\er)^*$. Since $\sim\!\!\b^*\sle
\b^*\!\!\sim$ and $\sim\!\!\er\sle\,\er$, we get $\a^*\sle
\,\sim\cup\, (\er)^*\!\!\a^*\cup\, \b^*\!\!\a^*$. Therefore,
$\a^*\sle\,\abes^*\sim$.\qed
\end{proof}


\begin{lemma}
\label{lem-equiv}
If $\cE$ is linear then the following propositions are equivalent:
$\a$ is confluent, $\abes$ is $\sim$-confluent, $\abes$ is
$\sim$-confluent on $\sim$-classes.
\end{lemma}

\begin{proof}
Since $\cE$ is linear, we have $\a^*\sle\,\abes^*\sim$ and
$\sim\abes^*\sle \abes^*\sim$. We prove that $\abes$ is
$\sim$-confluent if $\a$ is confluent: $\staseb\abes^* \,\sle \als\a^*
\,\sle \a^*\als \,\sle \abes^*\sim\sim\staseb$. We prove that $\a$ is
confluent if $\abes$ is $\sim$-confluent: $\als\a^* \,\sle\,
\sim\,^*\!\!\aseb\abes^*\!\!\sim \,\sle\, \sim\abes^*\sim\staseb\sim
\,\sle\, \abes^*\sim\sim\sim\staseb$. We now prove that $\abes$ is
$\sim$-confluent on $\sim$-classes if $\abes$ is $\sim$-confluent (the
inverse is trivial): $\staseb\sim\abes^* \,\sle \staseb\abes^*\sim
\,\sle \abes^*\sim\staseb\sim \,\sle \abes^*\sim\sim\staseb$.\qed
\end{proof}


\begin{theorem}
Type-checking is decidable if $\abes$ is weakly normalizing, $\cR$ is
finitely branching, $\abes$ is $\sim$-confluent on $\sim$-classes,
$\cE$ is linear and $\sim$ is decidable.
\end{theorem}

\begin{proof}
Type-checking is deciding whether a term $t$ has type $T$ in an
environment $\G$. A type for $t$ can be easily inferred. Then, one
checks that it is equivalent to $T$ (see \cite{coquand91book} for more
details). Thus, we are left to prove that $\ad^*$ is decidable. Since
$\cE$ is linear and $\abes$ is $\sim$-confluent on $\sim$-classes, by
Lemma \ref{lem-equiv}, $\a$ is confluent and $\ad^*=\ad$. Since $\cE$
is linear, by Lemma \ref{lem-ad}, $\ad\,=\abes^*\sim\staseb$. Since
$\abes$ is weakly normalizing and finitely branching ($\sim$-classes
are finite and $\b$ and $\cR$ are finitely branching), one can define
a function $nf$ computing a $\abes$-normal form of a term. We prove
that $t\ad^* u$ only if $nf(t)\sim nf(u)$ (the inverse is
trivial). Assume that $t\abes^* t'\sim u'\,^*\!\!\aseb u$. Since
$\abes$ is $\sim$-confluent on $\sim$-classes, $nf(t)\sim
nf(t')\,^*\!\!\aseb t'\sim u'\abes^* nf(u')\sim nf(u)$. Again, since
$\abes$ is $\sim$-confluent on $\sim$-classes, there exist $t''$ and
$u''$ such that $nf(t)\sim nf(t')\abes^* t''\sim u''\,^*\!\!\aseb
nf(u')\sim nf(u)$. Since $nf(t')$ and $nf(u')$ are $\abes$-normal
forms, we have $nf(t)\sim nf(u)$.\qed
\end{proof}


\newcommand{\Ri}{\o{R}}
\renewcommand{\sR}{\sim\!\!R}
\newcommand{\sRi}{\sim\!\Ri}
\newcommand{\Rs}{R\!\!\sim}
\newcommand{\Ris}{\Ri\!\sim}

\begin{lemma}
\label{lem-commut}
For all relation $R$, if $R$ $\sim$-commutes on $\sim$-classes then
$\sR$ is $\sim$-confluent on $\sim$-classes.
\end{lemma}

\newcommand{\oS}{\o{S}}
\newcommand{\oR}{\o{R}}

\begin{proof}
Let $S={\sim R}$. We prove that ${\oS^p\sim S^n}\sle {S^n\sim\oS^p}$
by induction on $n$.

\begin{lst}{\bu}
\item Case $n=0$. By induction on $p$. The case $p=0$ is
immediate. Case $p+1$: ${\oS^{p+1}\sim}= {\oS\oS^p\sim}\sle
{\oS\sim\oS^p}\sle {\sim\oS\oS^p}$ since ${\oS\sim}= {\oR\sim\sim}=
{\oR\sim}= \oS\sle {\sim\oS}$.

\item Case $n=1$. By induction on $p$.

\begin{lst}{--}
\item Case $p=0$. ${\sim S}= {\sim\sim R}= {\sim R}= {S\sle S\sim}$.
\item Case $p+1$. ${\oS^{p+1}\sim S}= {\oS\oS^p\sim S}\sle {\oS
S\sim\oS^p}\sle {S\sim\oS\oS^p}$ since ${\oS S\sim}= {\oR\sim\sim
R\sim}= {\oR\sim R\sim}\sle {R\sim\oR\sim}\sle {S\sim\oS}$.
\end{lst}

\item Case $n+1$. ${\oS^p\!\sim S^{n+1}}= {\oS^p\!\sim S S^n}\sle
{S\!\sim\oS^p S^n}\sle {S\!\sim\oS^p\!\sim S^n}\sle {S\!\sim
S^n\!\sim\oS^p}$ and we prove that ${S\sim S^n\sim}\sle {S^{n+1}\sim}$
by induction on $n$. The case $n=0$ is immediate. Case $n+1$: ${S\sim
S^{n+1}\sim}\sle {S\sim S^n\sim S\sim}\sle {S^{n+1}\sim S\sim}\sle
{S^{n+1}S\sim}$ since ${\sim S}= {\sim\sim R}= {\sim R}= S$.\qed
\end{lst}
\end{proof}


\begin{lemma}
\label{lem-sim-confl}
For all relation $R$, if $R$ is $\sim$-confluent on $\sim$-classes
then $\sR$ is $\sim$-confluent on $\sim$-classes.
\end{lemma}

\begin{proof}
If $R$ is $\sim$-confluent on $\sim$-classes then $R^*$
$\sim$-commutes on $\sim$-classes. Hence, by Lemma \ref{lem-commut},
$\sR^*$ is $\sim$-confluent on $\sim$-classes. Therefore, $\sR$ is
$\sim$-confluent on $\sim$-classes since $(\sR)^*\sle(\sR^*)^*$ and
$(\sR^*)^*\sle(\sR)^*\sim$.\qed
\end{proof}


\begin{theorem}
$\abes$ is $\sim$-confluent on $\sim$-classes if $\abes$ is strongly
normalizing, $\cE$ is linear, $\cRs$ is locally $\sim$-confluent and
$\cRs$ is locally $\sim$-coherent.
\end{theorem}

\begin{proof}
We first prove that $\b\cup\cR$ is $\sim$-confluent on $\sim$-classes.
In \cite{huet80acm}, Huet proves that a relation $R$ is
$\sim$-confluent on $\sim$-classes if $R\sim$ is strongly normalizing,
$R$ is locally $\sim$-confluent and $R$ is locally $\sim$-coherent. We
take $R=\b\cup\cRs$ and check the conditions. $R\!\!\sim$ is strongly
normalizing since $\abes$ is strongly normalizing and $\b$ and $\sim$
commute ($\cE$ is linear).  Local confluence: $\o\b\b\sle \b^*\o\b^*$
since $\b$ is locally confluent, $\o\cRs\b\sle \b^*\o\cRs^*\o\b^*$
after the proof of Lemma \ref{lem-beta-com}, and $\o\cRs\cRs\sle
\cRs^*\sim\o\cRs^*$ by assumption. Local coherence: $\cE\b\sle
\b\cE\sle \b\sim$ since $\cE$ is linear, and $\cE\cRs\sle
\cRs^*\sim\o\cRs^*$ by assumption.

So, $R=\b\cup\cR$ is $\sim$-confluent on $\sim$-classes. Therefore, by
Lemma \ref{lem-sim-confl}, $\sR$ is $\sim$-confluent on
$\sim$-classes. We now prove the theorem. We have $\abes^*\sle
(\sR)^*$ and $(\sR)^*\sle\abes^*\sim$ ($\b$ and $\sim$ commute since
$\cE$ is linear). Thus, $\staseb\sim\abes^*\sle
(\o\sR)^*\sim(\sR)^*\sle (\sR)^*\sim(\o\sR)^*\sle
\abes^*\sim\sim\staseb$.\qed
\end{proof}

Huet also proves in \cite{huet80acm} that $\cRs$ is locally
$\sim$-confluent iff its critical pairs are $\sim$-confluent, and that
$\cRs$ is locally $\sim$-coherent if $\cRs$ is left-linear and the
critical pairs between $\cRs$ and $\cE$ are $\sim$-confluent. So,
$\sim$-confluence is decidable whenever $\abes$ is strongly
normalizing, $\sim$ is decidable and $\cR\cup\cE$ is finite: it
amounts to checking whether the critical pairs between the rules, and
between the rules and the equations (in both directions), are
$\sim$-confluent.

Unfortunately, when considering type-level rewriting, confluence is
required for proving strong normalization. Whether strong
normalization can be proved by using local confluence only is an open
problem. Fortunately, confluence can be proved for a large class of
rewrite systems without using strong normalization, namely the
left-linear systems.


\begin{theorem}
$\abes$ is $\sim$-confluent on $\sim$-classes if $\cE$ is linear,
$\cRs$ is left-linear and $\cRs$ is $\sim$-confluent on
$\sim$-classes.
\end{theorem}

\begin{proof}
In \cite{oostrom94lfcs}, Van Oostrom and Van Raamsdonk prove that the
combination of two left-linear and confluent Combinatory Reduction
Systems (CRS) $\cH$ and $\cJ$ is confluent if all the critical pairs
between the rules of $\cH$ and the rules of $\cJ$ are trivial. We
prove the theorem by taking $\cH=\cRs\cup\cE$ and $\cJ=\b$, and by
proving that $\cH$ is confluent. Since $\cH^* \,\sle\, (\er)^*\sim$,
we have $\o\cH^*\cH^* \,\sle\, \sim(\o\er)^*(\er)^*\sim$. Since $\cRs$
is $\sim$-confluent on $\sim$-classes, by Lemma \ref{lem-sim-confl},
$\er$ is $\sim$-confluent on $\sim$-classes. Therefore,
$\sim(\o\er)^*(\er)^*\sim \,\sle\, \sim(\er)^*\sim(\o\er)^*\sim
\,\sle\, \cH^*\o\cH^*$.\qed
\end{proof}

Again, $\cRs$ is $\sim$-confluent on $\sim$-classes if $\er$ is
strongly normalizing and $\cRs$ is locally confluent and
$\sim$-coherent, which can be proved by analyzing the critical pairs
between the rules and between the rules and the equations (when $\cRs$
is left-linear) \cite{huet80acm}.


\section{Conclusion}

In \cite{blanqui01lics,blanqui01thesis}, we give general syntactic
conditions based on the notion of computability closure for proving
the strong normalization of $\b$-reduction and (higher-order)
rewriting. In this paper, we show that the notion of computability
closure can also be used for proving the strong normalization of
$\b$-reduction and (higher-order) rewriting modulo (higher-order)
equations. It is interesting to note that, in our approach, the
introduction of equations does not affect the conditions on rules:
although based on the same notion, equations and rules are dealt with
separately. Finally, one may wonder whether our method could be
extended to Jouannaud and Rubio's Higher-Order Recursive Path Ordering
(HORPO) \cite{jouannaud99lics,walukiewicz02jfp}, which also uses the
notion of computability closure for increasing its expressive power.\\

\small\noindent{\bf Acknowledgments.} I thank J.-P. Jouannaud, F. van
Raamsdonk and the referees for their useful comments on previous
versions of this paper. Part of this work was performed during my stay
at Cambridge (UK) thanks to a grant from the INRIA.


\end{document}